# Design of a "Digital Atlas Vme Electronics" ( DAVE ) Module

**Maurice Goodrick** [a], **Dave Robinson** [a], **Rick Shaw** [a],
**Martin Postranecky** [b*] , **Matthew Warren** [b]

[a] *Cavendish Laboratory, Department of Physics, University of Cambridge ,*
*J J Thomson Avenue, CB3 0HE, Cambridge, Great Britain*

[b] *Department of Physics and Astronomy, University College London,*
*Gower Street, London, WC1E 6BT, Great Britain*
*E-mail*: mp@hep.ucl.ac.uk

ABSTRACT: ATLAS-SCT has developed a new ATLAS trigger card, 'Digital Atlas Vme Electronics' ( "DAVE" ). The unit is designed to provide a versatile array of interface and logic resources, including a large FPGA. It interfaces to both VME bus and USB hosts.

DAVE aims to provide exact ATLAS CTP ( ATLAS Central Trigger Processor ) functionality, with random trigger, simple and complex deadtime, ECR ( Event Counter Reset ), BCR ( Bunch Counter Reset ) etc. being generated to give exactly the same conditions in standalone running as experienced in combined runs.

DAVE provides additional hardware and a large amount of free firmware resource to allow users to add or change functionality.

The combination of the large number of individually programmable inputs and outputs in various formats, with very large external RAM and other components all connected to the FPGA, also makes DAVE a powerful and versatile FPGA utility card.

KEYWORDS : Data acquisition circuits, Detector control systems, Digital electronic circuits

[*] Corresponding author.

## Contents

## 1. Introduction

ATLAS-SCT uses NIM electronics in USA15 ( ATLAS underground counting room ) to provide the trigger and veto logic needed in stand-alone physics runs. This needs to be updated as some of the components drift uncontrollably and adjustments need to be made manually.

### 1.1 Design and Development

Initially a purpose specific 6U VME card was proposed as a replacement, but this design quickly grew to incorporate many more generic features.

One significant enhancement was to adopt all ATLAS CTP functions to exactly duplicate ATLAS running conditions, whilst incorporating much generic functionality. This evolved the DAVE card into a powerful and flexible logic module, potentially useful to all ATLAS subsystems as well as for other non-ATLAS users [1], [2].

DAVE communication is by standard VME or by USB, allowing use inside VME-TTC crate or as a stand-alone on a bench-top. For this reason, +5V is the single power input ( from the VME backplane or a stand-alone power supply ), with all other power buses being generated on-board.

Firmware is developed in a modular fashion allowing code contributions from interested users. Initially the core SCT requirement ( vetoing trigger generation around a BCR ) was implemented.

Further development provides other CTP functionality, with random trigger generator up to 100kHz, simple and complex deadtime, busy gating, ECR, BCR, etc. generation to give exact same conditions in standalone as experienced in combined runs.

Firmware for this purpose is being copied from the CTP code to ensure identical operation.

Some other useful capabilities include BC/ORBIT source with fine-delay ( with 0.5nsec resolution ) for timing scans, generic counter facility, etc.

In addition, a large 'trigger sequence record / playback' is incorporated on a 72Mbit SRAM. This provides up to ~50 seconds history of trigger playback at 75 kHz L1 trigger rate ( e.g. on interrupt by system BUSY ).

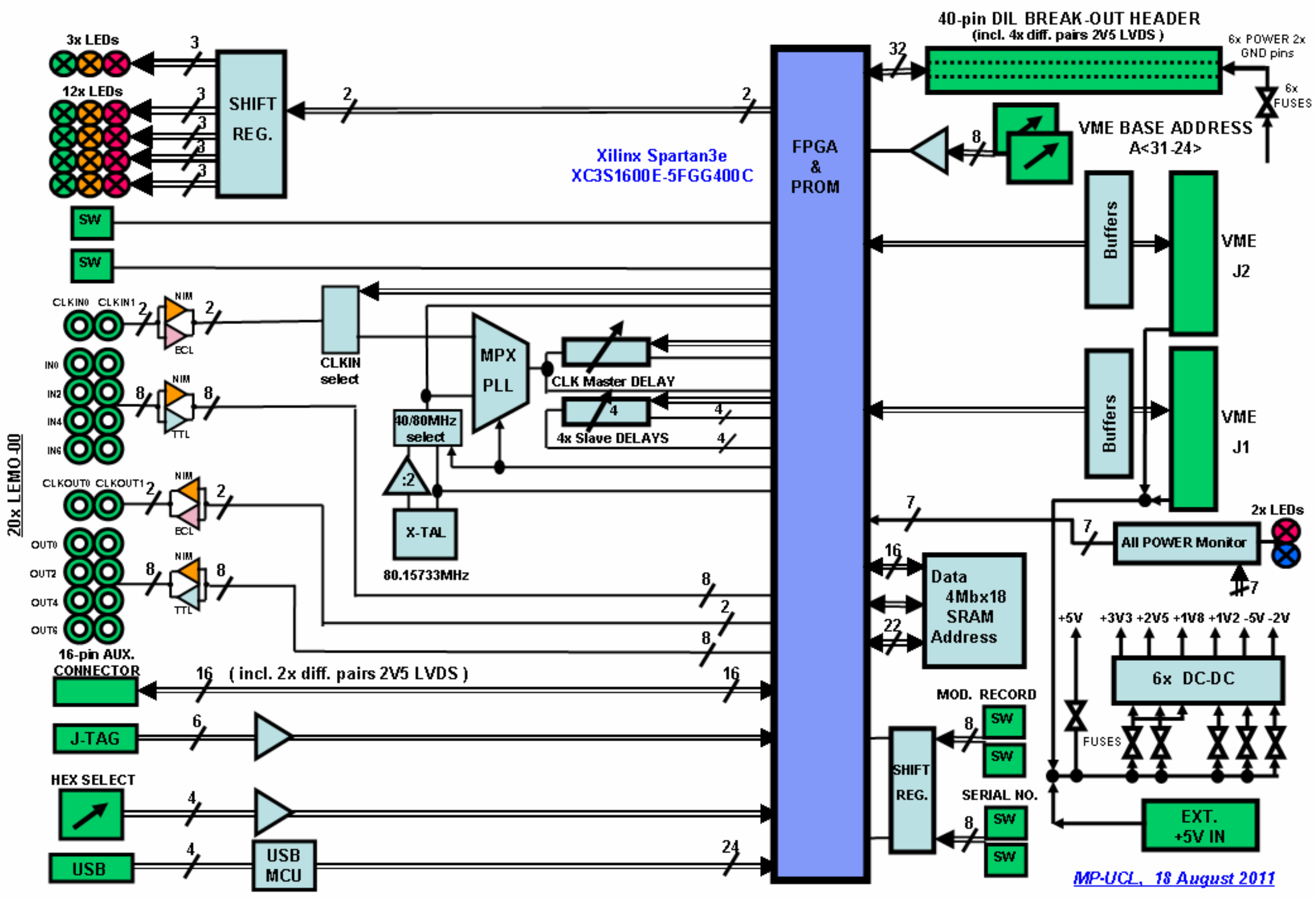

***Figure 1 :*** Block Diagram of DAVE-1 Hardware

## 2. Final Implementation

### 2.1 Hardware Description

The final design of DAVE is a single 6U standard VME-64 card with J1 and J2 VME connectors ( see ***Figure 1*** for block diagram of the hardware ) :

- Standard VME signals and pin assignments are used
- The VME slave interface supports A32/D16 access protocol
- The base address offset is address lines A31 – A24, selected by switches
- Address A0 is not used and always assumed to be zero
- VME access is D16 data transfers only, i.e. 16-bit words

- Address modifier codes for user and system data access are supported
- Block transfer, read-modify-write and address-only cycles are not yet supported

Additionally, the DAVE card has twenty separate LEMO 00 connectors on the front panel. These consist of two separate clock inputs and two clock outputs, all being individually switch-selectable as either –ve NIM or ECL standards, plus eight separate data inputs and eight data outputs, all also being switch-selectable as –ve NIM or TTL standards. All these inputs and outputs are individually programmable.

Further connectors accessible on the front panel are ( see ***Figure 2*** for the front panel connectors, displays and switches ) :

- 2mm/14-pin JTAG Header for programming the FPGA
- 2.54mm/10-pin DIL Header for USB Debug and download
- 4-pin USB port
- Auxiliary 2.54mm/14-pin DIL Header which has four programmable 2V5 LVDS I/O pins and twelve programmable 3V3 LVTTL I/O pins

There is also a separate daughter-card 2.54mm/40-pin DIL I/O Header on DAVE – also useable as a ribbon-cable connector – which has eight programmable 2V5 LVDS I/O pins, twenty-four programmable 3V3 LVTTL I/O pins and eight power and ground pins.

There is a rotary hex ‘Mode’ switch on the front panel which allows for selection of sixteen pre-programmed options of operation. There are also two reset switches, one with programmable functionality.

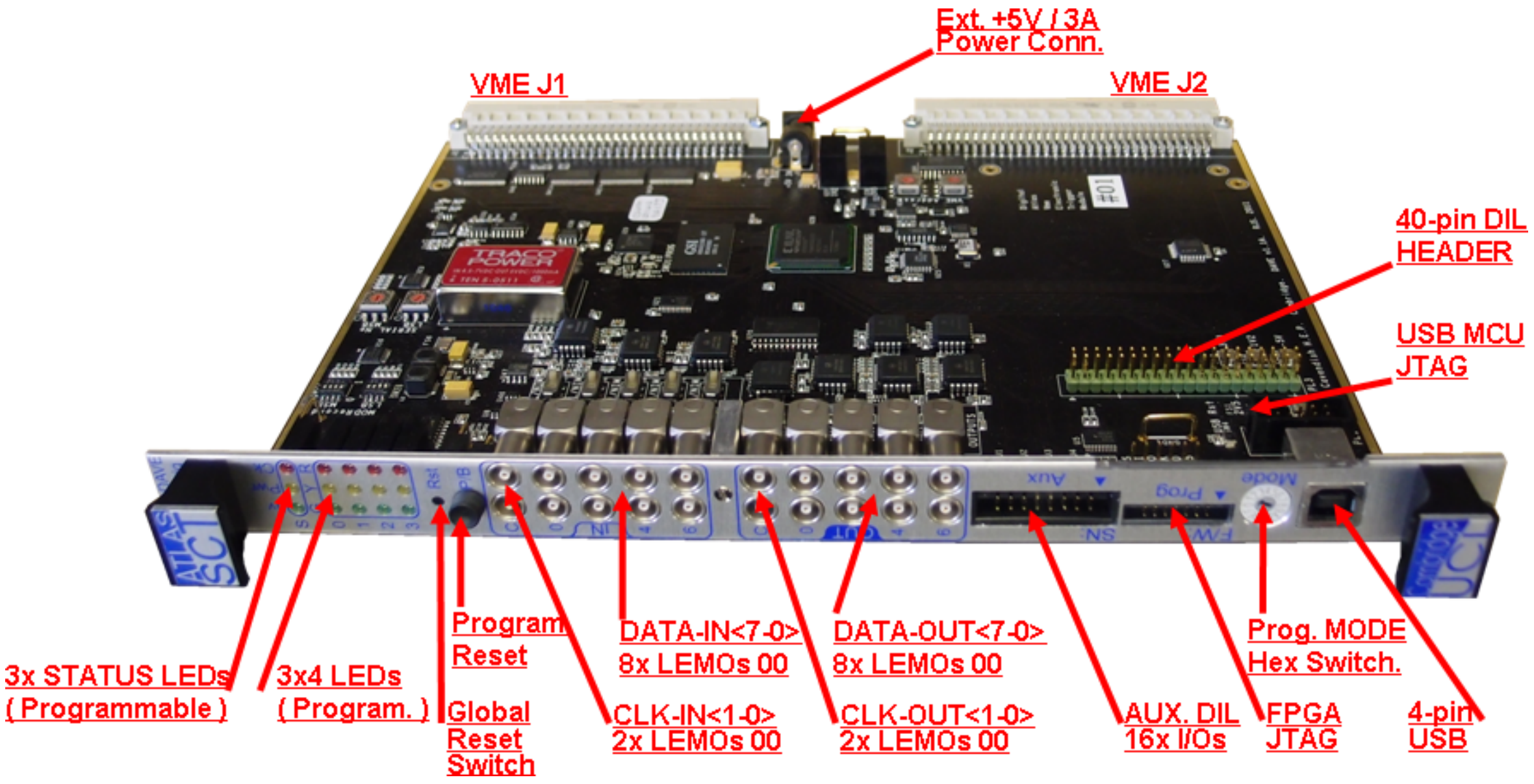

***Figure 2 :*** Front Panel Connectors, LED Displays and Switches of DAVE-1

Finally, there is a single 2.1mm Power Connector located between the J1 and J2 connectors, which is used to supply +5V DC / 2.5A power to DAVE when used as a stand-alone or bench-top card without any VME connections.

At the heart of DAVE sits a large Xilinx Spartan 3E 1600 ( XC3S1600E-FCG400C ) FPGA [3], together with Xilinx XCF8P PROM, providing all the functionality which may be required by users. Both FPGA and PROM can be programmed via the DIL JTAG connector on the front panel or via an auxiliary parallel SIL JTAG connector on the PCB.

In addition, there is a very large 4Mbx18 SRAM ( GS8642Z18GB 1671 ) [4] which can be used as a record or playback sequencer to aid any debugging or analysis of other subsystems. For example, up to ~50 seconds history of trigger information can be stored at 75 kHz L1 trigger rate. Sixteen DATA I/Os and twenty-two ADDRESS lines are connected to the FPGA. The SRAM can also be programmed using the JTAG chain.

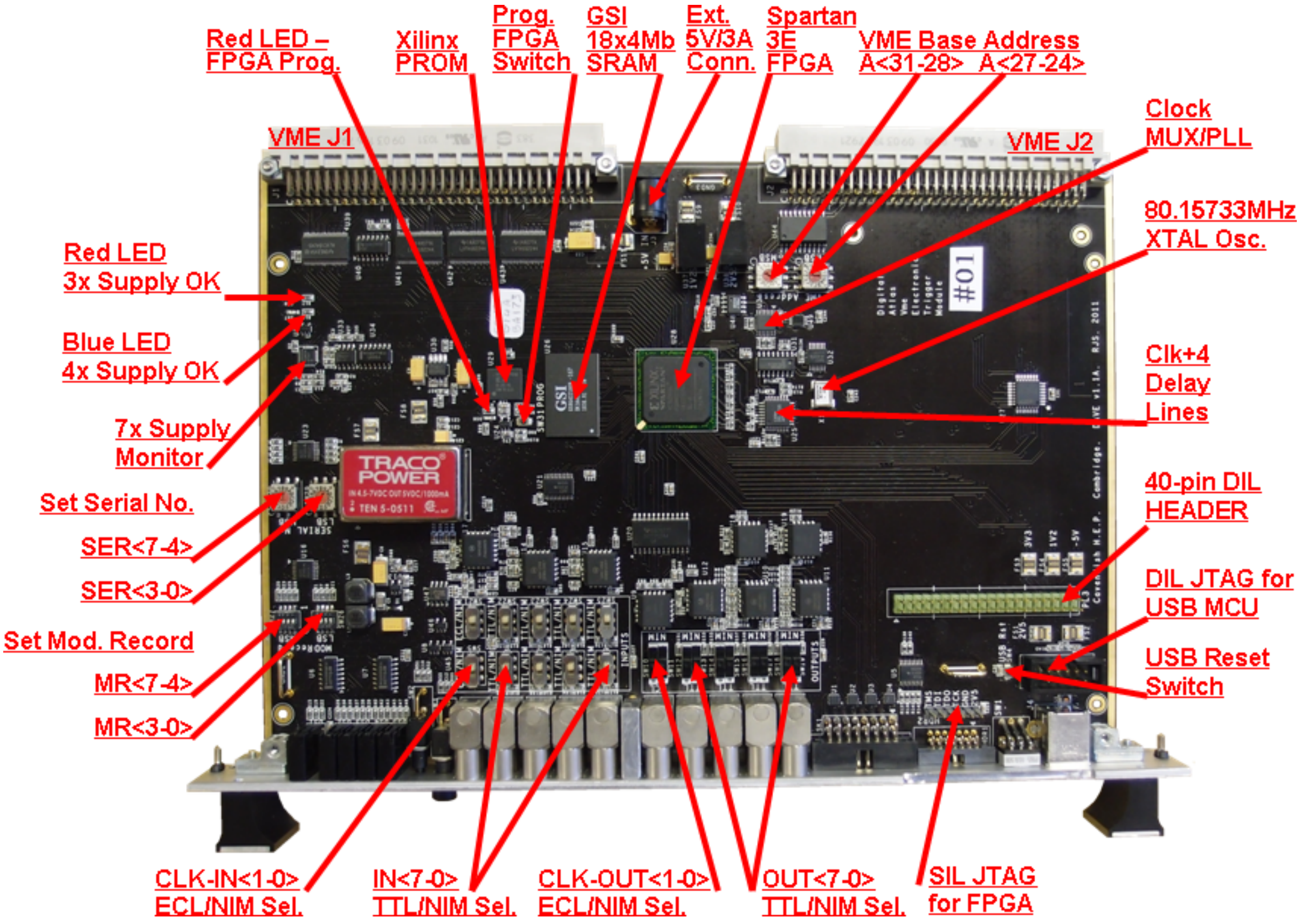

***Figure 3 :*** Layout of the major hardware components on the top of DAVE-1 PCB

To allow precise timing and programmable delays of clocks or data, a clock-plus-four channel programmable CERN-designed Delay Line 'Delay25' [5] is also connected to the FPGA. This allows for delays of up to 25nsec in steps of 0.5nsec of four programmable signals.

To provide a stand-alone capability, there is an on-board 80.15733MHz clock X-tal generator providing either ~40MHz or ~80MHz clock as selected via the FPGA. To ensure continuous, un-interruptible clock with either of the two individually selectable external clock inputs being used, the automatic clock multiplexer and PLL ( ICS581-02 ) [6] is used on DAVE. This provides automatic, smooth and glitch-less transfer to the stand-alone clock in case of the external clock failure.

Other hardware available on DAVE provides a fully-programmable fifteen LED display on the front panel, which can be used for status or diagnostic purposes, pre-settable switches for Serial and Modification Numbers and a full multi-supply voltage monitoring.

DAVE requires only a single +5V DC / 2.5A supply, together with GND, either via the VME backplane on J1 and J2 or via the stand-alone 2.1mm Power connector. All the other six power buses required ( +3V3, +2V5, +1V8, +1V2, -2V and -5V ) are generated by on-board DC-DC converters.

See ***Figure 3*** for the major hardware components on the top of DAVE-1 board.

Two prototype DAVE-P cards have been built and tested at Cambridge and at UCL early in 2011. A few assembly errors have been identified and corrected, as have one or two passive component changes. Overall, the prototype worked as intended. For example, numerous timing and jitter measurements were taken and were within the component specifications ( e.g. clock peak-to-peak jitter was below 200ps ).

Detailed description, further documentations and complete circuit and layout schematics of the DAVE card are available in the ‘DAVE User Guide’ [7], [8].

A further four DAVE-01 cards have been assembled and two are being used in firmware development and in actual SCT DAQ Electronics debugging at CERN.

## 2.2 Firmware Description

The firmware is still in development. The VME register map and control mechanisms have been written and are used in hardware testing both in England and at CERN.

A prototype version of the firmware is now available providing the core functions originally requested by the SCT ( vetoing trigger generation around a BCR ). The firmware controlling the operation of the SRAM has also been written and is capable of both recording and playing back trigger/BCR/ECR sequences. Having the large RAM is already proving very useful debugging tool.

See ***Figure 4*** for the diagram of the basic firmware blocks of DAVE-1 card. The current state of the firmware as available for DAVE is described, together with lists of the registers and commands required to use DAVE, in the ‘DAVE User Guide’ Firmware section [9].

In a normal combined physics running, the CTP ( Central Trigger Processor ) combines information from calorimeter and muon trigger and makes L1A ( Level-1 Accept ) decision on the basis of selected trigger menu. The CTP also sends an 8-bit Trigger-Type word with each L1A signal, via the TTC, to the front-end electronics of the various detectors including the SCT. The unique L1ID ( Level 1 Event Number ) identifying each event is periodically reset by the CTP issuing an ECR ( Event Counter Reset ) and the BCID ( Bunch Crossing Number ) is reset once per LHC orbit by the TTC system issuing BCR ( Bunch Counter Reset ). The CTP also receives Busy returns from the various detectors including the SCT.

For the SCT standalone runs, without the CTP information, all these signals and their simple and complex deadtime, busy gating, etc. must be generated on the DAVE to give exact the same conditions. Firmware for this purpose is being copied from the CTP code to ensure identical operation.

For detailed description of the CTP and SCT functionality see [10], [11].

For ATLAS-SCT DAQ Electronics Inter-Connections Schematics to CTP and TTC see [12].

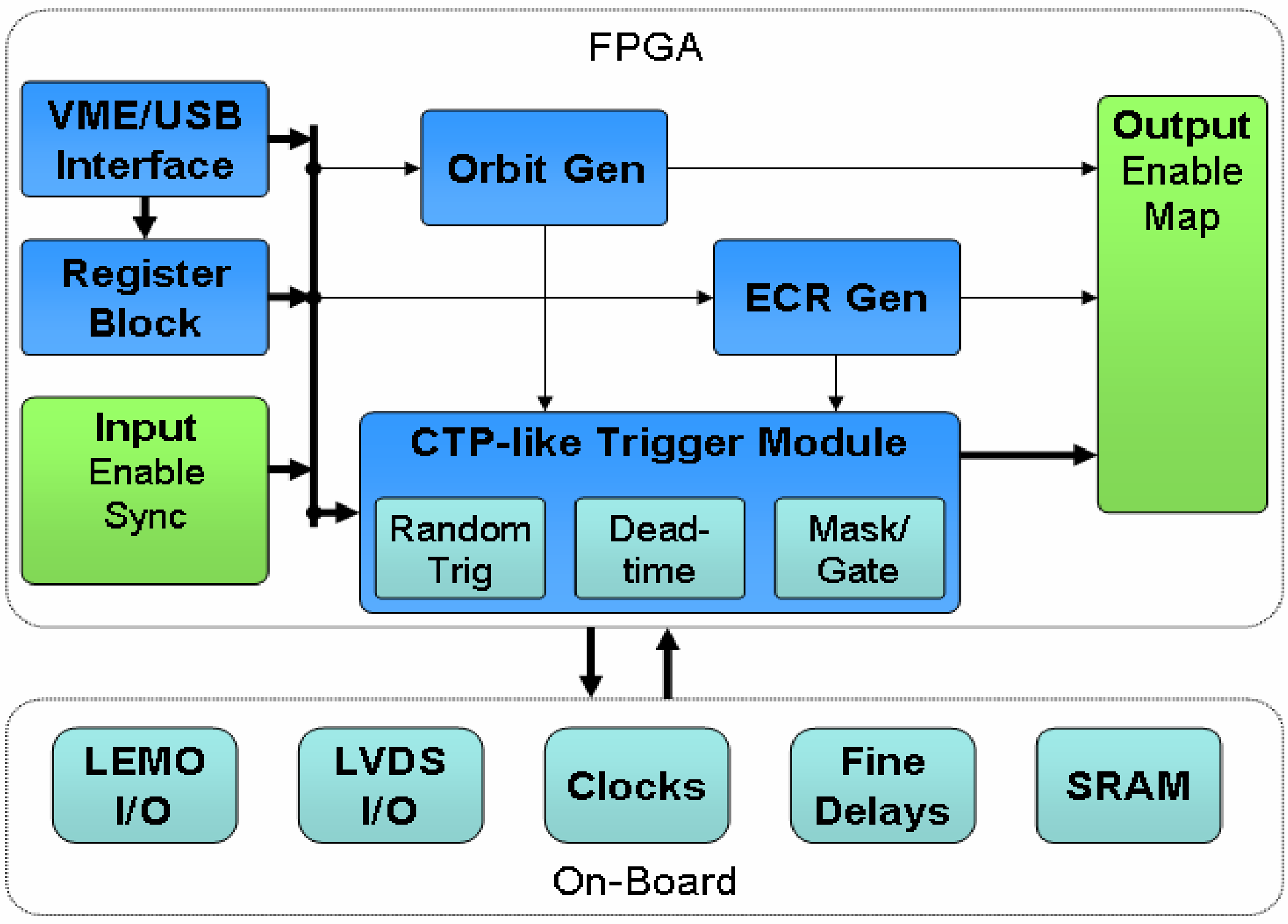

***Figure 4 :*** Basic Firmware Blocks of DAVE-1

Currently two DAVE modules are being used at CERN, one in SR1 ( ATLAS-SCT Test Electronics racks ) and the other in situ in the TTC crate in USA15 in the actual ATLAS-SCT DAQ Electronics racks. The fact that DAVE is actually in current use debugging SCT problems is already generating interest among other potential users.

## 3. Production and Use

A further 16 production modules are being manufactured, with some modifications based on the test results, to be delivered to users in the winter of 2011/2012. Additional firmware, suggested by various potential users, is also being developed.

## Acknowledgments

We acknowledge the help of various ATLAS colleagues and collaborators at CERN who helped to expand our plans for DAVE beyond the simple NIM replacement module and who provided additional user case suggestions for DAVE, especially Bruce Barnett, Carolina Gabaldone, Bruce Gallop, Andrej Gorisek, Stefan Haas, Iskander Ibragimov and Thilo Pauly.

Dave Robinson must be singled out as the ‘father’ of the idea to build this new module and thus giving the project its name.

We also acknowledge the financial support from the ATLAS collaboration to build these modules.